\documentclass[aps,prl,reprint,nobibnotes,
longbibliography,superscriptaddress]{revtex4-1}

\usepackage{t1enc}
\usepackage[utf8]{inputenc}
\usepackage[intlimits]{amsmath}
\usepackage{graphicx}
\usepackage{physics}

\begin{document}

\title{Josephson radiation and shot noise of a semiconductor
nanowire junction}

\author{David J.~van Woerkom}

\author{Alex Proutski}

\author{Ruben J.~J.~van Gulik}

\author{Tam\'{a}s Kriv\'{a}chy}

\affiliation{QuTech, Delft University of
Technology, 2600 GA Delft, The Netherlands}

\affiliation{Kavli Institute of Nanoscience, Delft University of
Technology, 2600 GA Delft, The Netherlands}

\author{Diana Car}

\affiliation{Department of Applied Physics, Eindhoven University of Technology,
5600 MB Eindhoven, The Netherlands}

\author{S\'{e}bastian R.~Plissard}

\affiliation{Department of Applied Physics, Eindhoven University of Technology,
5600 MB Eindhoven, The Netherlands}
\affiliation{Kavli Institute of Nanoscience, Delft University of
Technology, 2600 GA Delft, The Netherlands}

\author{Erik P.~A.~M.~Bakkers}

\affiliation{QuTech, Delft University of
Technology, 2600 GA Delft, The Netherlands}

\affiliation{Kavli Institute of Nanoscience, Delft University of
Technology, 2600 GA Delft, The Netherlands}

\affiliation{Department of Applied Physics, Eindhoven University of Technology,
5600 MB Eindhoven, The Netherlands}

\author{Leo P.~Kouwenhoven}

\author{Attila Geresdi}

\email[Corresponding author; e-mail: ]{a.geresdi@tudelft.nl}

\affiliation{QuTech, Delft University of
Technology, 2600 GA Delft, The Netherlands}

\affiliation{Kavli Institute of Nanoscience, Delft University of
Technology, 2600 GA Delft, The Netherlands}

\date{\today}

\begin{abstract}
We measured the Josephson radiation emitted by an InSb semiconductor
nanowire junction utilizing photon assisted quasiparticle tunneling in an
AC-coupled superconducting tunnel junction. We quantify the action of the local
microwave environment by evaluating the frequency dependence of the inelastic
Cooper-pair tunneling of the nanowire junction and find the zero frequency
impedance $Z(0)=492\,\Omega$ with a cutoff frequency of $f_0=33.1\,$GHz. We
extract a circuit coupling efficiency of $\eta\approx 0.1$ and a detector
quantum efficiency approaching unity in the high frequency limit. In addition to
the Josephson radiation, we identify a shot-noise contribution with a Fano
factor $F\approx1$, consistently with the presence of single electron
states in the nanowire channel.
\end{abstract}

\maketitle

The tunneling of Cooper pairs through a junction between two superconducting
condensates gives rise to a dissipationless current \cite{Josephon_1962} with a
maximum amplitude of the critical current, $I_c$ \cite{PhysRevLett.10.486}. Upon
applying a finite voltage bias $V$, the junction becomes an oscillating current
source
\begin{equation}
I_s(t)=I_c \sin(2\pi f t),
\end{equation}
with a frequency set by $hf=2eV$ where $h$ is the Planck constant and
$e$ is the electron charge. 

The Josephson radiation, defined by Eq.~(1) has mostly been investigated for
superconducting tunnel junctions \cite{Giaever_1965, Holst_1994, Deblock_2003},
metallic Cooper-pair transistors \cite{PhysRevLett.98.216802} and in circuit QED
geometries \cite{PhysRevLett.106.217005, PhysRevB.85.085435}. Recently, it has
also been proposed as a probe for topological superconductivity \cite{pikulin2012,
PhysRevLett.108.257001, PhysRevLett.111.046401}, which requires gateable
semiconductor Josephson junctions \cite{Doh_2006}.

In contrast to superconductor-insulator-superconductor (SIS) junctions,
Josephson junctions with a semiconductor channel feature conductive modes of
finite transmission probabilities \cite{xiang_2009, van2016microwave}, leading
to deviations from a sinusoidal current-phase relationship 
\cite{PhysRevLett.99.127005} and the universal ratio of the critical current and
the normal-state conductance \cite{PhysRevLett.10.486}.
Furthermore, soft-gap effects \cite{PhysRevLett.110.186803} have been shown to
result in excess quasiparticle current for subgap bias voltages, limiting
prospective applications such as topological circuits \cite{Mourik_2012} and
gate-controlled transmon qubits \cite{PhysRevLett.115.127002}.

Here we investigate the high-frequency radiation signatures of a voltage-biased
semiconductor Josephson junction \cite{Doh_2006} by directly measuring the
frequency-resolved spectral density. As a frequency-sensitive
detector, we utilize a SIS junction, where the photon-assisted tunneling current
\cite{Deblock_2003} is determined by the spectral density of the coupled
microwave radiation \cite{Tien_1963}. In addition to the detection of the
monochromatic Josephson radiation, we demonstrate the presence of a broadband
contribution, attributed to the shot noise of the nanowire junction
\cite{Blanter20001}, similarly to earlier experiments on carbon nanotube
quantum dots \cite{PhysRevLett.96.026803, PhysRevLett.108.046802}.

\begin{figure}
\centering
\includegraphics[width=0.5\textwidth]{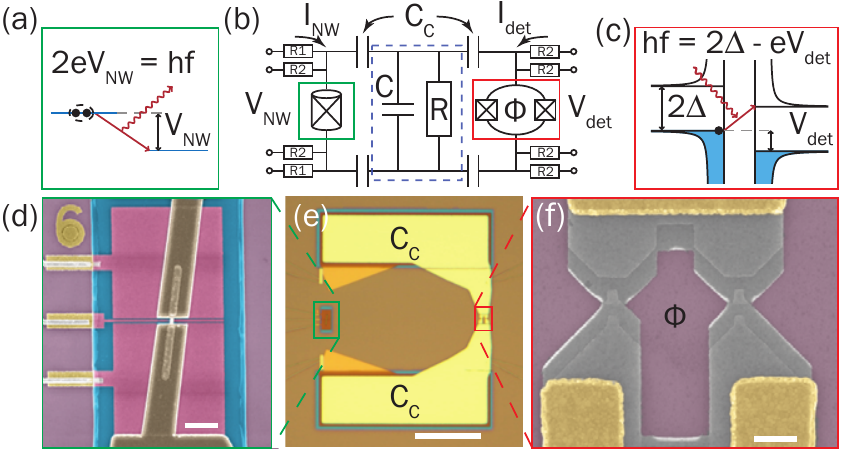}
\caption{(Color online) (a) Photon emission due to the inelastic Cooper-pair
tunneling between condensate levels shifted by the bias voltage,
$V_\textrm{NW}$. (b) The microwave equivalent circuit of the measurement setup,
where $R$ and $C$ in the blue dashed box represent the microwave losses and
stray capacitance, yielding a $2\pi f_0= (RC)^{-1}$ upper cutoff frequency.
The $C_c \gg C$ coupling capacitors have a negligible effect above a frequency
of $2\pi f_c= (RC_c)^{-1}$ with $f_c \ll f_0$, but allow for the application of
independent DC bias voltages $V_\textrm{NW}$ and $V_\textrm{det}$. The
$I_\textrm{NW}(V_\textrm{NW})$ and $I_\textrm{det}(V_\textrm{det})$
characteristics are measured through the Pt feedline resistors, depicted by
$R_1$ and $R_2$, respectively. (c) Photon-assisted quasiparticle tunneling for a
detector voltage bias $V_\textrm{det}$ and an incoming photon energy of $hf$.
(d) False colored scanning electron micrograph of the nanowire Josephson
junction contacted with NbTiN after being placed on three electrostatic gates.
(e) Bright field optical image of the coupling circuitry before the NbTiN
deposition step with the nanowire junction (green box) and the detector junction
(red box). (f) False colored micrograph of the detector split junction with an
applied magnetic flux $\Phi$. The scale bars depict $1\,\mu$m (d), $20\,\mu$m
(e) and $0.5\,\mu$m (f), respectively.}
\label{1}
\end{figure}

Our setup follows the geometry of earlier experiments utilizing SIS junctions
\cite{Deblock_2003}. In contrast, our microwave radiation source is an InSb
nanowire (NW) \cite{doi:10.1021/nl203846g} Josephson junction (Fig.~1d) with a
channel length of $100\,$nm. The junction leads (in brown in Fig.~1d) are
created by removing the surface oxides by Ar ion milling and then \emph{in-situ}
sputtering of NbTiN superconducting alloy. Owing to the highly transparent
contacts, this procedure enables induced superconductivity in the semiconductor
channel \cite{Mourik_2012, PhysRevLett.115.127002}. A predefined
gate structure (purple regions in Fig.~1d) provides electrostatic control of the
semiconductor channel and is covered by sputtering a $20\,$nm thick SiN$_x$
dielectric layer.

The $I(V)$ characteristics of the two junctions are measured in a standard four
point probe geometry via highly resistive Pt feedlines effectively decoupling
the on-chip elements (Fig.~1) thermally anchored at $20\,$mK from the
measurement setup. In order to gain access to a wider $V_\textrm{NW}$ range, we
use $R_1=1\,$k$\Omega$ in the nanowire biasing lines and $R_2=6\,$k$\Omega$ in
the voltage measurement leads (see Fig.~1b).

The detector SIS split junction is shown in Fig.~1f and is fabricated using
standard shadow evaporation techniques \cite{Dolan_1977}. The typical normal
state resistance was measured to be $20\,\textrm{k}\Omega$ for a nominal
junction area of $100\times100\,$nm$^2$.
The bottom and top Al layer thicknesses are $9$ and $11\,$nm, respectively. The
split junction geometry enables the flux control of the total Josephson coupling
of the detector. To measure the quasiparticle tunneling response, we set
$\Phi=\Phi_0/2$, with $\Phi_0=h/2e$ the flux quantum, to minimize the Josephson
coupling. We note that the minimal detector critical current is negligible
compared to that of the nanowire junction. Finally, we utilize two parallel plate capacitors of
$C_c\approx 400\,$fF with sputtered SiN$_x$ dielectric which couple the nanowire junction to
the detector in the frequencies of interest (Fig.~1e), yet enable independent
voltage biasing and current measurements in the DC domain.

\begin{figure}
\centering
\includegraphics[width=0.5\textwidth]{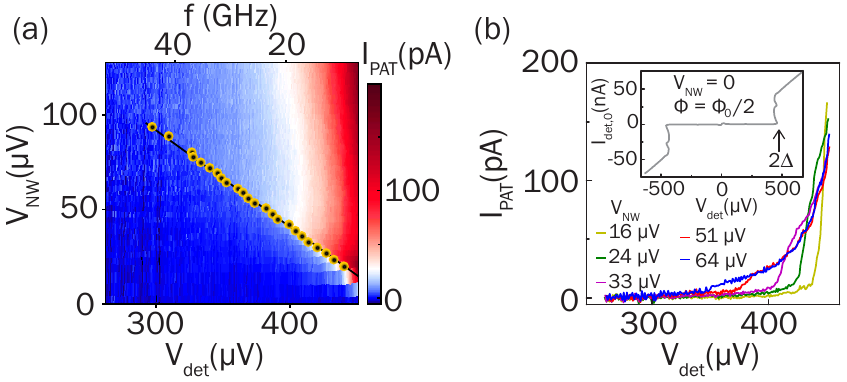}
\caption{(Color online) (a) Measured photon-assisted quasiparticle current
$I_\textrm{PAT}$ as a function of the detector bias voltage $V_\textrm{det}$ and
nanowire bias voltage $V_\textrm{NW}$. The orange dots denote the extracted
frequency on the upper axis for a given $V_\textrm{NW}$. The solid black line is
the best linear fit with $f/V_\textrm{NW}=475\,$MHz$/\mu$V. (b) Horizontal line
traces at different $V_\textrm{NW}$ values.
The inset shows the full $I_\textrm{det,0}(V_\textrm{det})$ characteristics of
the detector when the Josephson radiation is absent. Note the difference in the
current scale. The applied flux $\Phi=\Phi_0/2$ through the split junction
results in a suppressed detector supercurrent branch which minimizes its
Josephson radiation. The arrow depicts $2\Delta/e=480\,\mu$V, the onset of
the quasiparticle current.}
\label{2}
\end{figure}

The mesoscopic noise source under consideration is characterized by
its current noise density, $S_I(f)$ \cite{Blanter20001}, which results in
the voltage noise density  $S_V(f)=S_I(f) |Z(f)|^2$, where $Z(f)$ is the
complex frequency-dependent impedance of the coupling circuit. In Fig.~1b, we
depict a parallel $RC$ network resulting in $Z(f)=R(1-jf/f_0)/(1+f^2/f_0^2)$
with $2\pi f_0=(RC)^{-1}$ in the limit of negligible detector admittance,
$r_\textrm{det}^{-1}=dI_\textrm{det}/dV_\textrm{det} \ll R^{-1}$.

We deduce the voltage noise density $S_V(f)$ starting from the
equation for the photon-assisted current in the SIS detector
\cite{Tucker_1985,Deblock_2003}:

\begin{equation}
I_\textrm{PAT}(V_\textrm{det})=\int_0^\infty
\! S_V(f) \left(\frac{e}{hf}\right)^2
I_{QP,0}\left(V_\textrm{det}+\frac{hf}{e}\right) \dd{f},
\label{PAT}
\end{equation}
which describes the DC current contribution at an applied voltage
$V_\textrm{det}<2\Delta$. Crucially, this equation holds if the quasiparticle
current in the absence of radiation has a well-defined onset,
$I_{QP,0}(V_\textrm{det}<2\Delta)=0$ \cite{Deblock_2003} and in the limit of
weak coupling, where multiphoton processes do not contribute to the
quasiparticle current \cite{Tien_1963}. In addition, a detector with a sharp
quasiparticle current onset can reach the quantum limit \cite{Tucker_1985} where
each absorbed photon results in the tunneling of one quasiparticle. 

In the presence of a monochromatic radiation, where $S_V(f)\sim
\delta(f-\mathcal{F})$, Eq.~(2) describes the shift of the initial
$I_{QP,0}(V_\textrm{det})$ quasiparticle current by $\delta V_\textrm{det}=h
\mathcal{F}/e$. This is the case of the Josephson radiation \cite{Deblock_2003}
with $S_I(f)=\frac{I_c^2}{4}\delta(f-\mathcal{F})$, where
$h\mathcal{F}=2eV_\textrm{NW}$ with $V_\textrm{NW}$ the applied voltage bias on
the emitter junction with a critical current $I_c$. On the other hand, the
\emph{nonsymmetrized} quasiparticle shot noise is characterized by $S_I=eIF$
in the zero frequency and zero temperature limit with $I$ being the applied
current. The Fano factor, $F$ is characteristic to the mesoscopic
details of the junction \cite{Blanter20001}.

Note that Eq.~(2) can be handled as a convolution of $S_V(f)/(hf)^2$ and
$I_{QP,0}(V_\textrm{det})$. However, the inverse problem leading to $S_V(f)$ is
unstable due to the noise in the experimental data. To this end, we use Tikhonov
regularization \cite{doi:10.1080/03610926.2012.721916} to extract the noise
density measured by the detector (see \cite{rawdata} for details). It is to be
noted that the measured $I_\textrm{det,0}$ (see inset of the Fig.~2b) exhibits backbending due
to the self-heating effects in the leads of the superconducting tunnel junction,
therefore we used a monotonous $I_{QP,0}(V_\textrm{det})$ centered around the
same quasiparticle onset. However, the uncertainity of
$I_{QP,0}(V_\textrm{det})$ prevents the determination of the exact lineshape of
$S_V(f)$ which could indicate the linewidth of the
Josephson radiation \cite{PhysRevLett.22.1416}.

We demonstrate the detection of the Josephson radiation in Fig.~2. In panel (a),
we plot the PAT current contribution as a function of the DC bias voltages
$V_\textrm{det}$ and $V_\textrm{NW}$. In Fig.~2b, we show line traces
$I_\textrm{PAT}(V_\textrm{det})$ exhibiting well-defined onset values
corresponding to a monochromatic Josephson radiation tuned by $V_\textrm{NW}$.
Thus, we can extract the radiation frequency based on Eq.~(2) (orange dots in
Fig.~2a). By evaluating the relation between $V_\textrm{NW}$ and the radiation
frequency (black line in Fig.~2a), we find a ratio of $475\pm4.2
\frac{\textrm{MHz}}{\mu \textrm{V}}$ which is in reasonable agreement with
$\frac{2e}{h}\sim484\frac{\textrm{MHz}}{\mu \textrm{V}}$ expected for the case
of Cooper-pair tunneling \cite{PhysRevLett.18.287}. The intersect for $f=0$ is set by the
quasiparticle current onset to be $2\Delta/e=480\,\mu$V (see inset of
Fig.~2b).

The impedance $Z(f)$ of the environment results in a finite power dissipation
$I_c^2\textrm{Re}(Z(f))/2$  which gives rise to a DC current due to inelastic
Cooper-pair tunneling (ICPT) processes in the NW Josephson junction (see
Fig.~1a) \cite{Holst_1994}. This effect has been first addressed to calculate
the shape of the supercurrent branch in overdamped SIS junctions and purely
resistive environments \cite{ivanchenko1969}. Later, the theory was adapted for
high channel transmissions \cite{PhysRevLett.99.067008}. It has also been shown
that for an arbitrary $Z(f)\ll h/4e^2\approx6.5\,$k$\Omega$, the ICPT
contribution can be evaluated as \cite{Holst_1994}
\begin{equation}
I_\textrm{ICPT}=\frac{I_c^2 \textrm{Re}(Z(f))}{2V_\textrm{NW}},
\end{equation}
with a critical current $I_c$ and an applied voltage $V_\textrm{NW}$. Here,
the junction effectively probes the real component of the impedance $Z(f)$ at a
frequency $f=2eV_\textrm{NW}/h$.

In the following, we use a circuit model where the two independently measured
current values $I_\textrm{PAT}(V_\textrm{det})$ and
$I_\textrm{ICPT}(V_\textrm{NW})$ depend on the same microwave enviroment,
characterized by $Z(f)$. This model applies provided that the linear
resistance of the nanowire and the impedance of the detector, $r_\textrm{det}$,
are much higher than the effective shunt resistance of the circuit, depicted by $R$ in
Fig.~1b. In addition, the lumped element description of Fig.~1b is valid if the
circuit is much smaller than the characteristic wavelength $c/f \sim 1\,$mm. Our
structure, $~50\,\mu$m in size (see Fig.~1e), fulfills this condition. Note that
this is in contrast to a prior work \cite{PhysRevB.85.085435} where the
sample and detector were embedded in a transmission line resonator and thus the
effective impedance values were measured to be different.

\begin{figure}
\centering
\includegraphics[width=0.5\textwidth]{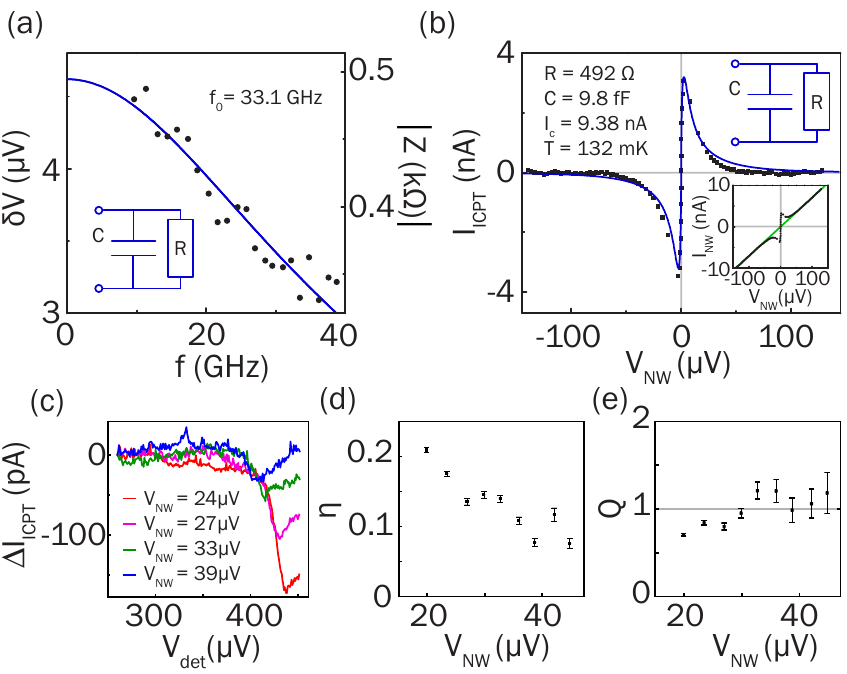}
\caption{(Color online) (a) The measured $\delta V(f)=I_c |Z(f)|$ voltage
fluctuation on the detector junction. The solid line depicts the fitted cutoff
with $f_0=(2\pi RC)^{-1}=33.1\,$GHz. Right vertical axis shows the impedance
$|Z(f)|$, see text. (b) Experimental $I_\textrm{ICPT}(V_\textrm{NW})$ trace of
the nanowire junction exhibiting a current peak due to the supercurrent branch.
The linear contribution with a resistance
$R_\textrm{NW}=14.03\,\textrm{k}\Omega$ (green solid line, see inset for raw
$I_\textrm{NW}(V_\textrm{NW})$ trace) is subtracted. The blue solid line depicts
the fitted curve with $I_c=9.38\,$nA critical current and a noise temperature
$T=132\,$mK.
(c) Variation of the nanowire junction current $\Delta I_\textrm{ICPT}$ as a
function of the detector voltage $V_\textrm{det}$. The extracted circuit
efficiency $\eta$ (d) and the detector quantum efficiency $Q$ (e) as a function
of $V_\textrm{NW}$, see text.}
\label{3}
\end{figure}

It is important to notice that the PAT current decreases with increasing
frequency (Fig.~2b). By correcting for the $\sim f^{-2}$ dependence in
Eq.~(2), we find that the fluctuation amplitude $\delta V = I_c |Z(f)| \sim
\sqrt{S_V}$ exhibits a characteristic cutoff frequency (Fig.~3a), even though
the current oscillation amplitude of the Josephson junction is constant, see
Eq.~(1).
Thus, we can attribute this cutoff to the coupling circuit impedance,
$Z(f)$. We find a good agreement between the experimental data and the
impedance of a single-pole $RC$ network (solid blue line in Fig.~3a) yielding to
a cutoff frequency $f_0=(2\pi RC)^{-1}=33.1\,$GHz.

Next, we turn to the measured $I(V)$ trace of the nanowire Josephson junction.
The inset of Fig.~3b shows the raw curve, which exhibits a supercurrent peak
around zero $V_\textrm{NW}$ and a linear branch. The latter fits to a linear
slope of $R_\textrm{NW}=14.03\,$k$\Omega$ (solid green line). We then extract
the $I_\textrm{ICPT}(V_\textrm{NW})$ component by subtracting this slope from
the raw measured data (black dots in Fig.~3b), which is an additive component to
the supercurrent peak unless the device has channels of transmission very close
to unity \cite{PhysRevLett.99.067008}. In order to find the critical current and
the noise temperature of the junction, we use the finite temperature solution of
Ivanchenko and Zil'bermann \cite{ivanchenko1969} with substituting $|Z(f)|$ as
the impedance of the environment \cite{rawdata}.
With this addition, we find an excellent agreement with the experimental data
(blue solid line in Fig.~3b), with $I_c=9.38\,$nA critical current. Notably,
with the now determined value of $I_c$, we can extract $R=492\,\Omega$ and
$C=9.8\,$fF fully characterizing the microwave environment of the junctions.
In addition, we find $I_c R_\textrm{NW}=132\,\mu$V, which indicates the induced
superconducting gap in the nanowire channel. This value is close to the induced
gap values measured earlier in similar devices \cite{Mourik_2012,
doi:10.1021/acs.nanolett.7b00540}. We also extract an effective noise
temperature $T=132\,$mK, which is higher than the substrate temperature of
$20\,$mK, similarly to earlier experiments \cite{PhysRevLett.99.067008}.

Thus far, we evaluated $I_\textrm{ICPT}(V_\textrm{NW})$ at
$V_\textrm{det}\approx 50\,\mu$V$\ll 2\Delta/e=480\,\mu$V, where
$I_\textrm{PAT}\approx 0$, thus the detector load is negligible. However,
depending on $V_\textrm{NW}$, we find a negative $\Delta
I_\textrm{ICPT}(V_\textrm{det})$, i.e.~a reduction of the emitter current, when
the detector threshold is on resonance with the emitted frequency (Fig.~3c). We
can understand this effect by the reduction of $Z(f)$ in Eq.~(3) in the presence
of a finite $r_\textrm{det}$ in parallel with $R$.
In first order, we find $\Delta
I_\textrm{ICPT}/I_\textrm{ICPT}=-\textrm{Re}({Z(f)})/r_\textrm{det}\approx
-R/r_\textrm{det}$. By using the measured DC current values, we evaluate the
efficiency of the coupling circuit to be the ratio of the absorbed and emitted
power $\eta=P_\textrm{det}/P_\textrm{emi}=2 I_\textrm{PAT}/I_\textrm{ICPT}$
(Fig.~3d). We find typical values spanning $0.1-0.2$, an order of magnitude
improvement over earlier reported values \cite{Deblock_2003,
PhysRevLett.96.136804}, however $\eta < 1$ owing to the resistive losses of the
device. Furthermore, the decrease of $\eta$ with increasing $f$ is consistent
with the low-pass nature of the coupling circuit. We also calculate the detector
quantum efficiency $Q=P_\textrm{det}/\Delta P_\textrm{emi}=2
I_\textrm{PAT}/\Delta I_\textrm{ICPT}$ (Fig.~3e) and find values scattering
around unity. This value directly measures the ratio of electron and photon rate
passing the detector junction, thus confirming that it is in the quantum limit
\cite{Tucker_1985}.

Finally, we note that the measured reduction $\Delta
I_\textrm{ICPT}/I_\textrm{ICPT} \ll 1$ directly confirms our initial assumption
of negligible detector load on the circuit. This proves that the analysis based
on a circuit model with the same $Z(f)$ for the nanowire junction and the SIS
detector is consistent.

\begin{figure}
\centering
\includegraphics[width=0.5\textwidth]{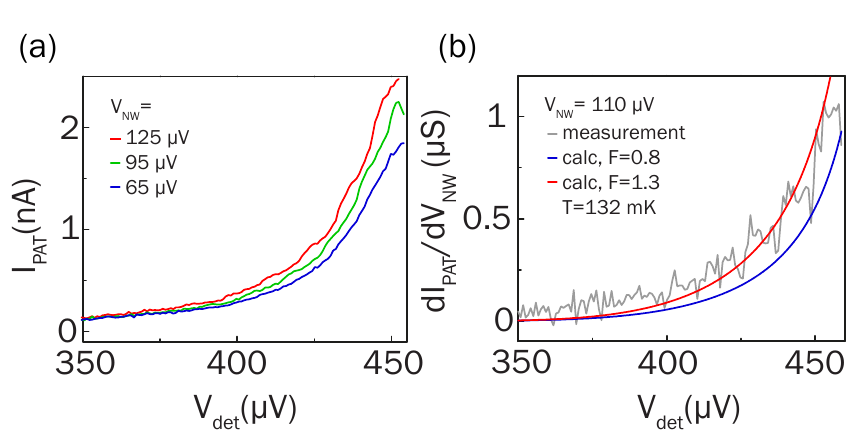}
\caption{(Color online) (a) Measured detector $I_\textrm{PAT}(V_\textrm{det})$
line traces at $V_\textrm{NW}=65$, $95$ and $125\,\mu$V bias voltage from the
bottom to top, respectively. (b) The measured $dI_\textrm{PAT}/dV_\textrm{NW}$
(light gray line) and the fitted curves at the top ($F=1.3$, red line) and the
bottom ($F=0.8$, blue line) of the confidence interval, respectively.}
\label{4}
\end{figure}

We now turn to the shot-noise contribution to $I_\textrm{PAT}$. We observe a
monotonous increase in $I_\textrm{PAT}$ with increasing $V_\textrm{NW}$ at any
$V_\textrm{det}$ consistently with the broadband $S_I$
(Fig.~4a). Note that, in contrast with the data shown in Fig.~2b, here the
contribution of the Josephson radiation is negligible. To quantify the shot-noise contribution,
we consider the derivative of the nonsymmetrized expression with respect to
$V_\textrm{NW}$ \cite{PhysRevLett.84.1986}:

\begin{equation}
\begin{split}
\frac{\dd S_I(f)}{\dd V_\textrm{NW}} = \frac{F}{R_\textrm{qp}}
\dv{V_\textrm{NW}}
\Bigl( & \frac{hf+eV_\textrm{NW}}{1-e^{-\beta(hf+eV_\textrm{NW})}}  \\
 & + \frac{hf-eV_\textrm{NW}}{1-e^{-\beta(hf-eV_\textrm{NW})}} \Bigr),
\end{split}
\end{equation}

where $\beta=1/k_\textrm{B}T$ is the inverse temperature \footnote{Note the we
omitted the voltage-independent terms in \cite{PhysRevLett.84.1986}}. We can
then calculate $\dd{I_\textrm{PAT}}/\dd{V_\textrm{NW}}$ by subsituting $\dd
S_I(f)/\dd V_\textrm{NW}$ in place of $S_I(f)$ in Eq.~(2). Using
the effective temperature $T=132\,$mK extracted earlier we find a confidence
interval of $F=0.8\ldots1.3$ (Fig.~4b). Considering that the channel length of
$100\,$nm is similar to the mean free path found earlier in the same nanowires
\cite{Gul_2012}, this result is consistent with ballistic transport
which is dominated by single electron channels of low transmission where $F=1$
\cite{dejong1997, Blanter20001}. In contrast, $F=1/3$ characteristic of
diffusive normal transport \cite{PhysRevB.46.1889} does not fit our data.

Furthermore, the measured $I_\textrm{NW}(V_\textrm{NW})$ and
$I_\textrm{PAT}(V_\textrm{NW})$ do not agree with a transport dominated by
multiple Andreev reflections, where a subgap structure is anticipated both in
the current \cite{PhysRevLett.78.3535} and in the shot noise
\cite{PhysRevLett.86.4104} depending on the channel transmissions. Our
experiment thus provides insight into the nature of the charge transport at
finite voltage bias in the nanowire Josephson junction and concludes that the finite
subgap current can be attributed to single electron states inside the induced
superconducting gap.

In conclusion, we built and characterized an on-chip microwave coupling
circuit to measure the microwave radiation spectrum of an InSb nanowire junction
with NbTiN bulk superconducting leads. Our results clearly demonstrate the
possibility of measuring the frequency of the Josephson radiation in a wide
frequency range, opening new avenues in investigating the $4\pi$-periodic
Josephson effect \cite{Lutchyn_2010} in the context of topological superconductivity
\cite{Oreg_2010}. Based on the Fano factor, the shot-noise contribution to the
measured signal demonstrates the presence of subgap quasiparticle states and
excludes multiple Andreev reflection as the source of subgap current of the
nanowire Josephson junction.

The authors acknowledge D.~Bouman, A.~Bruno, O.~Benningshof, M.~C.~Cassidy,
M.~Quintero-P\`{e}rez and R.~Schouten for technical assistance, and R.~Deblock
for fruitful discussions. This work has been supported by the Dutch Organization
for Fundamental Research on Matter (FOM), the Netherlands Organization for
Scientific Research (NWO) by a Veni grant, Microsoft Corporation Station Q, and
a Synergy Grant of the European Research Council.

\bibliographystyle{apsrev4-1}
\bibliography{references}

\end{document}